\documentclass[reprint]{revtex4}

\draft 

\usepackage{graphicx}
\usepackage{dcolumn}
\usepackage{bm}
\usepackage{color}

\begin{document}

\title{Transverse magneto-optical Kerr effect in subwavelength dielectric gratings} 

\author{Ivan S. Maksymov, Jessica Hutomo, and Mikhail Kostylev}

\affiliation{School of Physics, The University of Western Australia, 35 Stirling Highway, Crawley WA 6009, Australia}

\date{\today}

\begin{abstract}

We demonstrate theoretically a large transverse magneto-optical Kerr effect (TMOKE) in subwavelength gratings consisting of alternating magneto-insulating and nonmagnetic dielectric nanostripes. The reflectivity of the grating reaches $96\%$ at the frequencies corresponding to the maximum of the TMOKE response. The combination of a large TMOKE response and high reflectivity is important for applications in $3$D imaging, magneto-optical data storage, and magnonics.

\end{abstract}

\maketitle 

\section{Introduction}

The transverse magneto-optical (MO) Kerr effect (TMOKE) is defined as a change of reflected intensity of \textit{p}-polarised light when the direction of the external static magnetic field is changed from the saturated state $+M_{\rm{s}}$ to $-M_{\rm{s}}$, being $M_{\rm{s}}$ the saturation magnetisation \cite{Zve97}. The attainment of a large TMOKE response is important for many practical applications that include, but not limited to, $3$D imaging \cite{Aos10}, magnonics \cite{Kru10,Ser10}, and MO data storage \cite{Zve97}.

The natural TMOKE response of continuous single magnetic thin-films is weak and its detection is challenging. Consequently, a large and growing body of research investigates different mechanisms of the enhancement of the TMOKE in thin-film multilayers \cite{Arm13}, magneto-plasmonic crystals \cite{Wur08,Bel11,Che12,Kos13,Bel13,Hal13}, and nanoantennas \cite{Che11,Bon11}. All nanostructures mentioned above consist solely of ferromagnetic and nonmagnetic (e.g., gold) metals or contain a nonmagnetic metal nanostructure combined with a magneto-insulating layer.

Pure ferromagnetic magneto-plasmonic nanostructures made, e.g., of nickel (Ni) or permalloy (Py=Ni$_{\rm{80}}$Fe$_{\rm{20}}$) \cite{Che12,Kos13,Che11,Bon11} are gaining increasing attention due to their potential for integration with microwave magnonics devices (magnons are the quanta of magnetisation precession in ferromagnetic media). For instance, pure ferromagnetic nanostructures -- magnonic crystals -- are used in magnonics for the manipulation of spin waves \cite{Kru10,Ser10}. The standard techniques with which to read-out information carried by spin waves are MO Kerr effect (MOKE) and Brillouin light scattering (BLS) spectroscopy. However, the natural interaction between light and dynamic magnetisation is weak because the frequency of the precession of magnetisation ($1-–50$ GHz) and the velocity of spin waves ($\sim3 \times 10^3$ m/s) are widely disparate from those of light ($450–-590$ THz and $\sim3 \times 10^8$ m/s). Thus, the MOKE and BLS spectroscopy will benefit from the increase of the TMOKE response.

\begin{figure}[htb]
\centering\includegraphics[width=8cm]{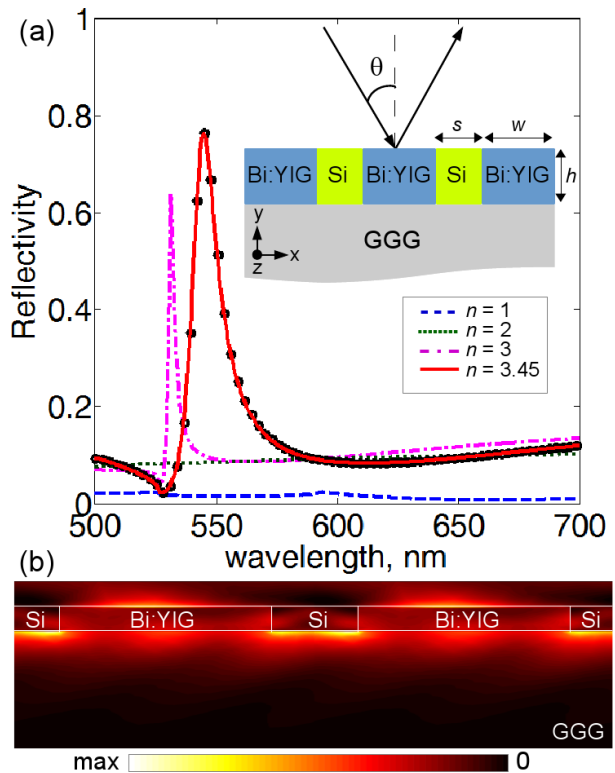}
\caption{(a) Inset: Schematic of the proposed grating. The length of the rulings of the grating along the \textit{z}-axis is assumed to be infinite, $h = 100$ nm, $w = 264$ nm, and $s = 113$ nm. Main panel: Reflectivity spectra of the grating calculated for different values of the refractive index $n$ of the material filling the grooves between the Bi:YIG nanostripes. The grating is illuminated by \textit{p}-polarised light incident at angle $\theta = 35^{\rm{o}}$. Lines –- in-house Rigorous Coupled-Wave Analysis (RCWA) software. Dots –- commercial CST Microwave Studio software for $n = 3.45$. (b) The total intensity profile in the plane of incidence at $\theta=35^{\rm{o}}$ at the frequency corresponding to the maximum of reflectivity for $n = 3.45$.}
\end{figure}

The enhancement of the TMOKE response is challenging not only due to the weak interaction of light with magnetic matter. In magneto-plasmonic nanostructures the enhanced TMOKE response is compromised by a relatively low reflectivity of $\sim20-25\%$ (see, e.g., \cite{Che12,Hal13}). Furthermore, in our work \cite{Kos13} we have shown that the maximum of the TMOKE response occurs at the frequency of the Fano resonance. Due to the asymmetry of the line shape of Fano resonances the maximum of the TMOKE response is often shifted from the maximum of the reflectivity towards its minimum.

In this paper, we demonstrate theoretically a large TMOKE response in subwavelength gratings consisting of alternating magneto-insulating and nonmagnetic dielectric nanostripes. Without loss of generality, as the magneto-insulating material we employ yttrium iron garnet doped with bismuth (Bi:YIG). As a dielectric material we choose silicon (Si). Whereas Bi:YIG is known to exhibit large MO activity, the high-refractive-index Si nanostripes help to achieve high reflectivity. Also, absorption losses in Bi:YIG and Si are very low as compared with nonmagnetic and ferromagnetic metals. As a result, the proposed grating offers a large TMOKE response accompanied by a remarkably high ($\sim96\%$) resonant reflectivity.

It is noteworthy that other iron garnets with a comparable refractive index, such as YIG that has been used in magnonics for decades \cite{Ser10}, can be employed to achieve the main result of this work. Thus, our grating can also serve as a magnonic crystal with simultaneously optimal magnetic and optical properties essential for the enhancement of the MOKE and BLS spectroscopy.

As the grating period is subwavelength, only the zero-order reflected and transmitted light is present in the far field, with higher-order diffraction modes cut off. The high reflectivity exhibited by subwavelength gratings depends on the angle of incidence and it is attributed to the excitation of a waveguide mode through phase matching by the grating. The waveguide mode then recouples back (again through grating phase matching) to radiative modes and produces the observed high reflection \cite{new_theory,Wan93,Mat04,Pet04,Mar04,Bis06}. The recoupling occurs at the frequencies corresponding to the position of the resonant Wood's anomaly, which in turn appears in the spectral range when the guided mode co-exists with the allowed diffraction modes \cite{new_theory,Wan93}. As also shown in \cite{Wan93}, the resonance spectral range is linked to the Rayleigh wavelength \cite{new_theory} at the given angle of incidence. Furthermore, the line shape of the resonance is asymmetric due to Fano interference \cite{new_theory,Mir10}. 

The complex resonance behaviour of the proposed subwavelength dielectric grating is a mechanism that differs it from conventional MO diffraction gratings having a large period as compared with the wavelength of incident light \cite{Sou98}. Moreover, similar to magneto-plasmonic gratings, the reflectivity of conventional MO gratings is low. The potential of subwavelength dielectric nanostructures to enhance the MO Kerr effect has been confirmed recently in \cite{Mar14}. By using an idealised analytical model of a subwavelength dielectric grating suspended in air, it has been shown that the polar and longitudinal configurations of the MO Kerr effect \cite{Zve97}, which are different from the TMOKE, can be enhanced without losing in reflectivity.

\section{Results and Discussion}

The inset in Fig. $1$(a) shows the schematic of the proposed subwavelength dielectric grating and defines its dimensions along with the polarisation of light incident at angle $\theta$. The grating consists of alternating nanostripes made of yttrium iron garnet doped with bismuth (Bi:YIG) and silicon (Si). The Bi:YIG--Si periodic structure sits on top of a gadolinium gallium garnet (GGG) layers that needs to be used as a seed substrate for the growth of realistic Bi:YIG nanostructures (see, e.g., \cite{Ame14}). The typical refractive index of Bi:YIG and GGG is $n_{\rm{Bi:YIG}}=2.2+0.0011i$ and $n_{\rm{GGG}}=1.97$, correspondingly \cite{Poh13}. Because the refractive index contrast between Bi:YIG and GGG is low, in the absence of the Si nanostripes our in-house RCWA software \cite{Moh95} predicts the reflectivity of the grating of just $2\%$ at $\theta=35^{\rm{o}}$ (Fig. $1$). This result shows the crucial and adverse impact of the GGG substrate on optical properties of realistic iron garnet nanostructures. 

However, at $n = 3.45$ corresponding to Si, we observe that the reflectivity reaches $\sim78\%$. This result is confirmed by CST Microwave Studio software implementing a Finite Integration Technique. Fig. $1$(b) shows the intensity profile in the plane of incidence at $\theta=35^{\rm{o}}$ at the frequency corresponding to the maximum of reflectivity of the Bi:YIG--Si grating.
	
Because our RCWA software \cite{Moh95} cannot simulate gyrotropic materials we use CST Microwave Studio to investigate the enhancement of the TMOKE associated with the resonance of the Bi:YIG--Si grating. In the TMOKE configuration the static magnetic field is perpendicular to the plane of incidence. We use the conventional expression \cite{Zve97} to quantify the TMOKE response: $\delta=R(+M_{\rm{s}})-R(-–M_{\rm{s}})$, where $R$ denotes the reflectivity. We do not normalise $\delta$ to the reflectivity $R(0)$ because in many practical applications (e.g., in magnonics \cite{Kru10}) one measures the difference between $R(+M_{\rm{s}})$ and $R(-–M_{\rm{s}})$ without knowing $R(0)$. We also avoid artificial enhancement of the TMOKE that can appear when $R \rightarrow 0$. The reflectivity $R$ changes due to the magnetic field induced change of the boundary conditions at the surface of the Bi:YIG nanostripes of the grating \cite{Zve97,Bel11}. In simulations, the dielectric permittivity tensor $\epsilon(M_{\rm{s}})$ has the following nonzero components: $\epsilon_{\rm{11}}$ = $\epsilon_{\rm{22}}$ = $\epsilon_{\rm{33}}$ = $n^{2}_{\rm{Bi:YIG}}$, $\epsilon_{\rm{13}} = ig$, and $\epsilon_{\rm{31}} = -ig$, where $g$ is the value of the gyration. For a saturated magnetisation $g_{\rm{Bi:YIG}}=0.005$ \cite{Poh13} . 

The solid line in Fig. $2$(a) shows the reflectivity spectrum of the Bi:YIG--Si grating. By changing the sign of the gyration in the dielectric permittivity tensor we simulate the change from $+M_{\rm{s}}$ to $-M_{\rm{s}}$. The solid line in Fig. $2$(b) shows the corresponding TMOKE response $\delta$. We compare these results with the reflectivity and  TMOKE response of a continuous Bi:YIG film of the thickness $h$ situated on top of the GGG substrate (the dashed lines in both panels of Fig. $2$). We observe a two orders of magnitude enhancement of the TMOKE response of the grating as compared with the continuous film.  

\begin{figure}[htb]
\centering\includegraphics[width=11cm]{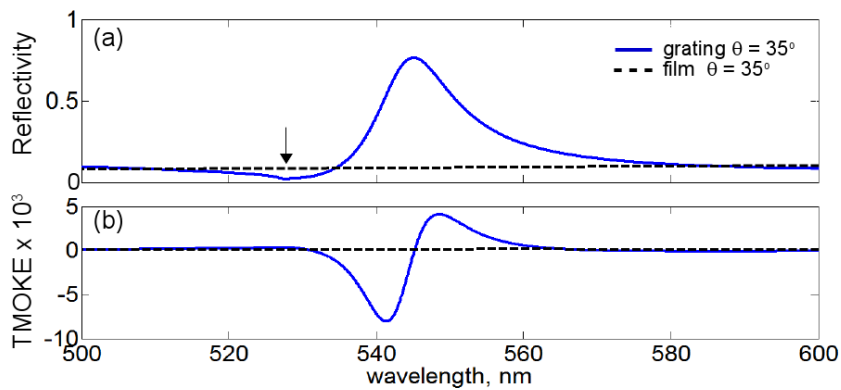}
\caption{Reflectivity spectrum (a) and the TMOKE response (b) of the continuous Bi:YIG film (dashed lines) and Bi:YIG--Si grating (solid lines) at $\theta=35^{\rm{o}}$. The Rayleigh wavelength is indicated by the arrow in (a).} 
\end{figure}

The resonance in the reflectivity spectrum of the grating is due to the excitation of a waveguide mode and its recoupling back to the incident region. Both processes occur through phase matching by the grating and they are correlated with the Rayleigh wavelength and diffractions modes of the grating \cite{Wan93}. The Rayleigh wavelength corresponds to the Wood's anomaly, which is due to one of the spectral orders appearing or disappearing in the case of the diffracted wave lying in the plane of the grating. Although the Wood's anomalies are usually associated with the resonant excitation of surface plasmons \cite{Che12}, they are also present whenever there exists a surface mode. In agreement with a theory of periodic dielectric waveguides \cite{Pen75}, our simulations show that our grating supports a surface wave propagating along the Bi:YIG--Si nanostructure [Fig. $1$(b)]. One can see that the intensity of the surface wave is higher in the high refractive index Si regions, which additionally confirms the crucial role of the Si nanostripes.  

Fig. $3$(a) shows that the sharp resonance peaks in the reflectivity spectra of the grating appear at the spectral range corresponding to the $+1$st order reflected/transmitted diffraction modes shown in Fig. $3$(b). We note that the same result was predicted in Ref.~\cite{Wan93}. The spectral position of the resonance peaks is linked to the Rayleigh wavelength $\lambda_{\rm{R}} = \Lambda (sin \theta \pm n_{\rm{i}}) / \pm m$, where $\Lambda$ is the grating period, $n_{\rm{i}}$ the refractive index of the sub- and superstrate of the grating, and $m$ is the diffraction order. Consequently, the resonance peaks shift as a function of $\theta$ [Fig. $3$(a)]. It is noteworthy that at $\theta=35^{\rm{o}}$ the grating supports both reflected and transmitted diffraction modes. At the smaller $\theta$ the grating supports only the transmitted diffraction modes.

The co-existence of the guided and diffraction modes in a narrow spectral range creates conditions for Fano interference, which leads to the formation of the asymmetric line shape [Fig. $2$(a)]. Another characteristic feature of the Fano resonance -- the suppression of the reflectivity (or antiresonance) \cite{Mir10} -- is also present in the spectrum and it occurs at the Rayleigh wavelength [the arrow in Fig. $2$(a)].

\begin{figure}[htb]
\centering\includegraphics[width=11cm]{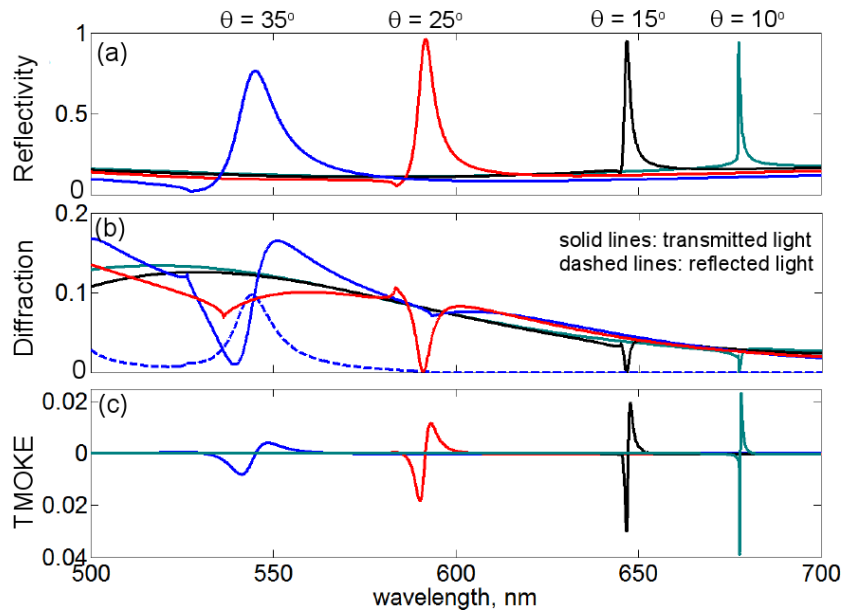}
\caption{Reflectivity spectra (a), diffraction efficiency (b) and the TMOKE response (c) of the subwavelength dielectric grating for different angles of incidence $\theta$ (blue line: $\theta=35^{\rm{o}}$, red line: $\theta=25^{\rm{o}}$, black line: $\theta=15^{\rm{o}}$, sea green line: $\theta=10^{\rm{o}}$). In (b), the solid (dashed) line denotes the diffraction efficiency of transmitted (reflected) light. The minor secondary peaks in the spectra occur at the Rayleigh wavelength.}
\end{figure}

In Fig. $3$(a) we also observe that the linewidth of the resonance peaks gets narrower. As a result, the reflectivity increases from $78\%$ to remarkably high $96\%$. The direct implication of the linewidth narrowing is a drastic increase in the TMOKE response that reaches considerable $\delta=-0.04$ at $\theta=10^{\rm{o}}$ [Fig. $3$(c)]. Indeed, in accord with the expression quantifying the TMOKE response $\delta$, a shift in the position of a narrow resonance peaks gives rise to a higher TMOKE response than the same shift in the position of a broad peak.

Finally, we refer to Fig. $3$(a) in our work \cite{Kos13} where we investigate a Py grating of the same period as the subwavelength dielectric grating. However, instead of Si the Py grating has air grooves. One can see that at $\theta=35^{\rm{o}}$ the TMOKE response of the Py grating is $3.2$ times lower as compared with that of the dielectric grating. Furthermore, the reflectivity of the Py grating is $<20\%$ at the frequency corresponding to the maximum of the TMOKE response. 

Under the assumption that Si nanostripes are lossless, our simulations predict that at the resonance the absorption losses in the Bi:YIG--Si grating constitute $\leq5\%$. In the off-resonance regime the losses are $\sim0.1\%$. Even though in reality there will be always more losses due to the effect of the substrate and possible fabrication imperfection, the proposed grating will always outperform magneto-plasmonic gratings in terms of absorption losses. For instance, at the resonance the losses in the aforementioned Py grating constitute $\sim75\%$. For further comparison, in a nonmagnetic gold grating of the same geometry the losses are $\sim25\%$.  

\section{Conclusions}

We have proposed and verified numerically an efficient mechanism of the enhancement of the TMOKE in subwavelength dielectric gratings. We have demonstrated a large TMOKE response accompanied by a very high reflectivity. This result is not readily attainable using magneto-plasmonic gratings. We envision the application of the proposed grating in a variety of devices for high-definition imaging, magneto-optical data storage, and magnonics.  
\\
\\
\noindent This work was supported by the UWA UPRF scheme and Australian Research Council. We thank Prof. S. Samarin (UWA School of Physics) for valuable comments and suggestions.


\end{document}